\documentclass[pra,aps,showpacs,floatfix,twocolumn]{revtex4-2}
\usepackage{color}
\usepackage{colordvi}
\usepackage{graphicx,amssymb,epsfig,amsmath}
\usepackage{epstopdf}
\usepackage{amsmath}
\usepackage{bm}
\usepackage{amsfonts}

\usepackage{amsmath}
\usepackage{amssymb}
\usepackage{amsthm} 
\usepackage{bm}
\usepackage{amsfonts}
\usepackage{color}

\begin{document}

%%paper title
%%For line breaks \\ can be used within title 

\title{Exploring vortex formation in rotating Bose-Einstein condensates beyond mean-field regime}
\author{Budhaditya Chatterjee}
%\email{budhaditya.e11237@cumail.in}
\affiliation{ Department of Physics, Chandigarh University, Gharuan,  India}

%%escape two column mode for title, affiliation and abstract
%%by giving \twocolumn command as shown

\begin{abstract}
	
  The production of quantized vortices having diverse structures is a remarkable effect of rotating Bose-Einstein condensates. Vortex formation described by the mean-field theory is valid only in the regime of weak interactions. The exploration of the rich and diverse physics of strongly interacting BEC requires a more general approach. This study explores the vortex formation of strongly interacting and rapidly rotating BEC from a general ab initio many-body perspective. We demonstrate that the quantized vortices form various structures that emerge from an intricate interplay between the angular momentum and many-body interaction. We examine the distinct impact of the angular velocity and interaction energy on the vortex formation. Our analysis shows that, while the angular rotation generally augments the vortex formation, the interactions can enhance as well as impede the vortices production.
  
\end{abstract}

\maketitle

%%insert keywords separated by comma using \keywords{words}
%\keywords{Bose-Einstein Condensate, Vortex lattice, Numerical simulation}

%%include \pacs{number} to print the PACS number
%\pacs{03.75.Lm; 67.85.-d; 05.30.Jp}

%%close the twocolumn escape here

%%include \doinum{number}for the DOI number in the header
%%include \volnum{number} for the volume number in the header
%%include \year{yyyy} for  year of publication in the header
%%include \pgrange{num--num} page range of article in the header
%%include \artcitid{num} for the article citation id
%%include \lp to print last page of the article
%%include \setcounter{page}{pagenum} for the exact starting page of the article

%\doinum{}
%\artcitid{}
%\volnum{}
%\year{}
%\pgrange{0}
%\setcounter{page}{0}

\section{Introduction}
The formation of quantized vortices is a natural consequence of angular momentum in rotating fluids. These quantized vortices can be produced and observed in  Bose-Einstein condensate (BEC) setups \cite{Abo-Shaeer01,Coddington2003, Donadello14,Wilson15}. Various techniques are used to generate quantized vortices in BECs, such as laser stirring \cite{Madison00}, rapid rotation \cite{Hodby01,Haljan01}, direct phase-defect imprinting \cite{Leanhardt02}.

Vortex signatures in BEC manifest themselves in density nodes and phase discontinuity, observed in the density or momentum distribution. A variety of fascinating phenomena has been theoretically predicted and experimentally observed, such as vortex bending \cite{Ripoll01} Tkachenko oscillations, quantum Hall states, vortex lattice formation \cite{Aftalion2003a}, vortex amalgamation\cite{Engels2003a} and melting \cite{Fischer2003a,Coddington2003}.    

A rotating BEC is generally analyzed from a mean-field perspective by solving the Gross–Pitaevskii equation. The mean-field framework, which assumes that all the particles are coherent and condensed in a single state, can aptly predict a variety of phenomena, including the formation of a vortex lattice \cite{Tkachenko,Tsubota02}.  

The mean-field theory necessitates a system with a huge number of particles having negligible interactions and becomes inexact when the quantum correlations become significant. In particular, both for extremely strong interactions and rapid rotations, fragmentation in BECs occurs, i.e., multiple orbitals become macroscopically occupied \cite{Mueller}. Hence, the fundamental assumptions of the mean-field method are no longer fulfilled, and the corresponding solutions become inaccurate. Therefore, it is necessary to employ a general quantum many-body framework to account for fragmentation and quantum correlations appropriately in such situations.

The multi-configuration time-dependent Hartree method for bosons (MCTDHB) ~\cite{alon08} is a suitable general quantum many-body method capable of addressing the strong interaction regime \cite{Lode20,Lode_mctdhx}. The MCTDHB method has been applied to compute the static and dynamics properties in a variety of strongly interacting and fragmented system  \cite{chatterjee15,cao17,Mistakidis18a,Streltsova14,sakmann16,lode17,tsatsos17,chatterjee17a,chatterjee17b,chatterjee20}.

This work explores the vortex formation in a strongly interacting, rapidly rotating BEC from a general ab initio quantum many-body perspective by numerically solving the Schr\"odinger equation using the MCTDHB method. Primarily through the analysis of the one-body density, we identify the characteristic vortex structure signatures that emerge as the interaction coupling and angular velocity are varied. Our work identifies the distinct impact of angular momentum and interaction energy on vortex formation and its structure.  

The paper is structured as follows. In Section~\ref{sec:model} we discuss the model Hamiltonian and the setup of our system.

In Section~\ref{sec:method}, we describe our numerical method. In Section~\ref{sec:results}, we display and discuss the results. Section~\ref{sec:conclusions} concludes the paper.

\section{Model} \label{sec:model}
We consider a two-dimensional rotating BEC comprising $N$ atoms, trapped in a quasi-two-dimensional circular disk potential.
The many-body Hamiltonian in the co-rotating frame is given by
\begin{equation}
\begin{aligned}
H= {} -\sum_{i=1}^{N}\frac{\hbar^{2}}{2M}\nabla_{i}^{2}+\sum_{i=1}^{N}V_{pot}(\bm{r}_{i}) + \sum_{i=1}^{N} V_{rot}(x_i,y_i) \\ +\sum_{i<j}V_{int}(\bm{r}_{i}-\bm{r}_{j}).
\label{Eq.Ham}
\end{aligned}
\end{equation}
Here, $N$ is the number of atoms each of mass $M$. The first term represents the kinetic energy of the setup.
The trapping potential $V_{pot}$ is a hard-wall circular disk potential in the $X-Y$ plane modeled as
\begin{equation}
V_{pot}(r) = \left\{
\begin{array}{ll}
0 & r \leq a \\
1000 & r > a
\end{array}
\right. ,
\label{eq:hard_sphere_pot}
\end{equation}
In the ultracold temperature regime, the energies of the atoms are low and consequently only the s-wave scattering is relevant. Therefore, the exact (Van-der wall-like) potential between the atoms can be modeled as a contact delta-like interaction between two atoms at $\bm{r}_{i}$ and $\bm{r}_{j}$ as $V_{int}(\bm{r}_{i}-\bm{r}_{j}) = g\delta(\bm{r}_{i}-\bm{r}_{j})$.
Here, $g$ is the interaction coupling constant and regulates the strength of the interaction.
However, owing to the computational convergence problems pertaining to the usage of actual delta potentials in two-dimension \cite{doganov}, we model the delta function as a narrow normalized Gaussian function as
\begin{equation}
\delta(\bm{r = \bm{r}_{i}-\bm{r}_{j}}) = \frac{\exp\left(-\frac{r^{2}}{2\sigma^{2}}\right)}{2 \pi \sigma^2}, \label{Vcontact}
\end{equation}
The width of the Gaussian is taken as $\sigma =0.25$, chosen to correspond with the physics of the true delta potential.
The BEC rotates in the $z$ direction with an angular velocity $\Omega$. The corresponding Hamiltonian term is given as $V_{rot} = \Omega \bm{L_z}$, where $\bm{L_z} = -i\hbar(x\partial_{y} - y\partial_{x})$.
The ground-state of the co-rotating Hamiltonian \Ref{Eq.Ham} provides the steady-state solution of the system.
For our numerical computation, we rescale the Hamiltonian \Ref{Eq.Ham} by $\frac{\hbar^2}{2ML^2}$ where $L$ is a typical length-scale of the system. This makes all quantities in the rescaled Hamiltonian dimensionless and defines the units of computed quantities.

\section{Method} \label{sec:method}
We use the numerically exact multi-configuration time-dependent Hartree method for bosons (MCTDHB) implemented in the MCTDH-X package~\cite{Lode_mctdhx} to solve the many-body Schr\"odinger equation.
In this method, the many-body wavefunction is expanded in terms permanents
\begin{equation}
\vert \Psi \rangle = \sum_{\vec{n}} C_{\vec{n}} (t) \vert \vec{n}; t \rangle. \label{Ansatz}
\end{equation}
The permanents are constructed as a linear combination of time-dependant, symmetrized, variationally optimal bosonic basis states as:
\begin{equation}
\vert \vec{n} ; t \rangle = \frac{1}{\sqrt{\prod_{\alpha=1}^M n_\alpha !}} \prod_{\alpha=1}^M \left( b_\alpha^\dagger(t) \right)^{n_\alpha} \vert \textrm{vac} \rangle,
\end{equation}
Here, $N$ bosons are distributed in $M$ single-particle states.
Subsequently, a time-dependent variational method is used \cite{Kramer} to generate two sets of equations of motion ~\cite{alon08} which are solved to obtain the many-body solutions. Fundamentally a time-dependent method, the ground-state is obtained using propagation in imaginary time.

\section{Results} \label{sec:results}

We focus on the strong interaction and rapid rotation regime wherein the particles are no longer entirely condensed, and hence the mean-field approximation is no longer accurate.
We consider $N = 100$ in the disk potential. To explore the formation of quantized vortices, we analyze the one-body density as a function of the interaction strength $g$ and the angular velocity $\Omega$.

\begin{figure}[h]
\centering
\includegraphics[width=0.9\columnwidth,keepaspectratio]{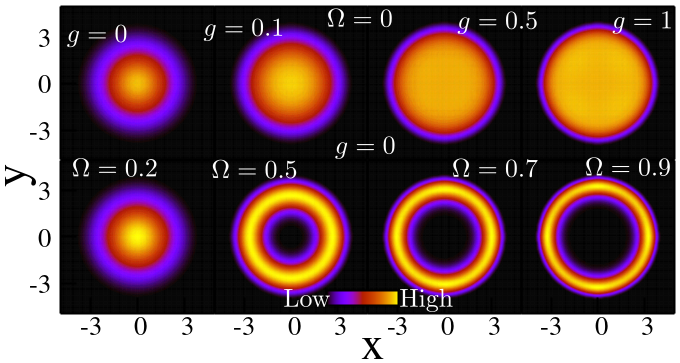}
\caption{(color online) Top panel: One-body densities for different interaction strengths for no rotation ($\Omega = 0$). Bottom panel: For different rotational velocity with no interaction ($g=0$) 
\label{cap:norot}}
\end{figure}
We aim to examine the impact of angular velocity and interaction strength on vortex formation. To that end, we first study the non-rotating $\Omega = 0$ and non-interaction case $g=0$ as a reference. Figure \ref{cap:norot} (top panel) depicts the one-body densities for no rotation $\Omega = 0$.   Absent both interaction and rotation, the density shows a layered profile -- the density is largest at the center and reduces along the radius being lowest at the rim. This can be attributed to the dominating influence of the kinetic energy and the hard-wall potential forcing the particles towards the center.

As interaction increases ($g = 0.1$), the repulsive interaction results in the particles being pushed out, countering the kinetic energy influence. As the interaction becomes significant $g = 0.5--1$, the interaction energy dominates, resulting in a uniform disk density.

Figure \ref{cap:norot} (bottom panel) depicts the one-body densities for no interaction $g = 0$.
For significant angular velocity $\Omega \geq 0.5$, an extensive region of very negligible density is formed at the center. This can be attributed to the centrifugal effect of the rotation pushing the particles radially outward. The size of this region increases with increasing angular velocity and, at $\Omega = 0.9$, covers most of the disk.

\begin{figure}[h]
\centering
\includegraphics[width=0.9\columnwidth,keepaspectratio]{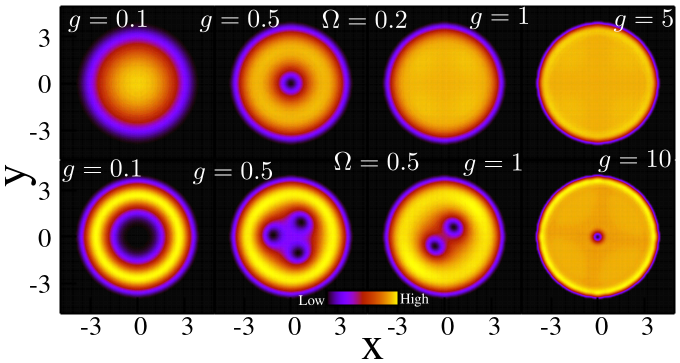}
\caption{(color online):One-body densities for different interaction strengths for low angular velocity ($\Omega = 0.2-0.5$). \\
Top panel: $\Omega = 0.2$ Bottom panel: $\Omega = 0.5$ 
\label{cap:lowrot}}
\end{figure}
We now introduce both rotation and interaction in our system. First, we investigate the physics with low angular velocity ($\Omega = 0.2, 0.5$).

The introduction of interaction and rotation leads to their interplay, impacting the vortex production.

Fig \ref{cap:lowrot}(top panel) depicts the one-body density for $\Omega = 0.2$. For interaction strength $g = 0.1$,the one-body density is roughly similar to the non-interacting case -- a layered profile. For $g = 0.5$, the rotation results in a vortex formation exactly at the center of the disk. This is revealed as a density node or a ‘hole’ in the one-body density. The density assumes a uniform pattern in the other regions. The vortex disappears for stronger interaction $g>1$, and we acquire a roughly uniform density profile.

Fig \ref{cap:lowrot}(bottom panel) depicts the one-body density for moderate angular velocity $\Omega = 0.5$. The higher angular velocity enhances the vortex production in the BEC.

Here, for $g=0.1$, we observe the formation of a large density hole at the center covering about one-third the radius of the disk. The formation of this density hole can be attributed to the dual effect of rotation and interaction pushing the particles to the rim.

As interaction increases to $g=0.5$, we observe the formation of vortex structures. The large central hole transforms into a low-density triangular region with a vortex at the vertices. This triangular triple-vortex pattern transforms into a double-vortex structure for $g=1$. For very strong interaction $g=10$, we observe a solitary vortex exactly at the center. This single vortex persists for stronger interactions. We observe that the higher angular velocity enhances the vortex production, especially for the intermediate interaction regime. However, very strong interaction reduces the number of vortices produced.

\begin{figure}[h]
\centering
\includegraphics[width=0.9\columnwidth,keepaspectratio]{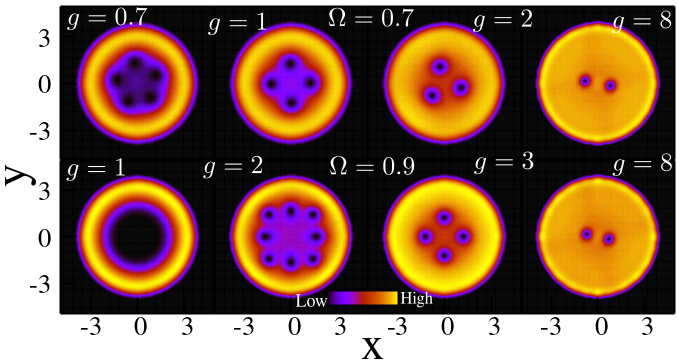}
\caption{(color online):One-body densities for different interaction strength for high rotational velocity ($\Omega = 0.5-0.9$). \\
Top panel: $\Omega = 0.7$. Bottom panel: $\Omega = 0.9$ 
\label{cap:highrot}}
\end{figure}

We now explore the larger angular velocity regime. Fig \ref{cap:highrot}(top panel) depicts the one-body density for $\Omega = 0.7$. Because of the larger angular velocity, the interplay between the angular momentum and interaction occurs at a much lower interaction coupling. For small interactions, we have a similar central region hole as before (not shown here).

Increasing the interaction coupled with the faster rotation leads to an enhancement of vortex production and the richness of various structures formed. At $g=0.7$, the central circular hole transforms into a pentagonal shape of low density. The vortexes arrange themselves into the vertices of this pentagonal geometry. The pentagonal vortex geometry transforms into a diamond-shaped configuration for $g=1$ with vortexes at the four vertices. Further interaction $g=2$ reduces the structure into a triangular geometry comprising 3 vortexes. Finally, at strong interaction $g=8$, we obtain a two-vortex structure that persists for stronger interaction, demonstrating the suppression of vortex formation with very high interactions.

Fig \ref{cap:highrot}(bottom panel) depicts the one-body density for $\Omega = 0.9$. Here, the rapid rotation results in a large hole region for $g=1$ -- the rapid rotation effectively destroys the previously obtained pentagonal and diamond vortex lattices. At $g=2$ instead, we obtain a square structure comprising 8 vortexes. Interestingly, the eight vortexes do not arrange as an octagon but a square -- this specific geometrical configuration minimizes the energy. Further interaction $g=3$ reduces the 8-fold vortex structure into a 4-vortex diamond. The consistent reduction of vortex formation for higher interactions is also seen here. At $g=8$, the structure is reduced into two vortexes, which persist for larger interactions. 

These simulations display the contrasting influence of the interaction strength compared to the rotational velocity. While the increasing angular velocity $\Omega$ leads to an increased formation of vortexes having a rich plethora of structures, the influence of interaction strength $g$ is more involved. Whereas increasing from small to intermediate interaction enhances the vortex production, very strong interactions destroy the vortex structures, leading to a reduction of the vortexes.  

\section{Conclusions} \label{sec:conclusions}
We explored the vortex formation in a rotating Bose-Einstein condensate through an ab initio numerical simulation of the many-body Schr\"odinger equation. The vortex structure and formation depend on the interplay between the angular velocity and the interaction strength. 

Only a single vortex is generated for slow rotation and intermediate interaction strength. The vortex production is enhanced for faster angular velocity, and various vortex structures form. Here, the impact of interaction becomes critical --  increasing the interaction to intermediate strength assists in the vortex structure formation. However, very strong interaction strength results in suppressing the vortex formation and the melting of the corresponding structures.

For intermediate rotation ($\Omega = 0.5$), we observe a transformation of a large central hole into a triangular vortex geometry when the interaction increases. For faster rotations ($\Omega = 0.7$), richer structures emerge for intermediate interaction strength ($g = 0.7-3$), that includes a pentagon, diamond,  triangle geometry finally melting into a double vortex. For high-speed rotations ($\Omega = 0.9$), we observe an octet square structure transforming into a triangle and a double vortex as the interaction increases. The simulation demonstrates the vortex formation and destruction tendency of interactions. The number and structures of vortices are most significant for intermediate interactions and decrease as the interaction increases. 

In general, we observe that while both interaction strength and angular velocity facilitate the formation of the quantized vortex of various structures, a powerful interaction leads to the breakup of the vortex formation. The higher rotational rate favors the formation of a greater number of vortices having various structures, which is especially evident in the intermediate interaction regime.

%%Use section* for acknowledgements
\section*{Acknowledgement}
We thank A.U.J. Lode and the other contributors for the development and support of the MCTDHX package.  

%%use \balance somewhere in the left column of the last page to balance the two columns in the end page

%%References section


\begin{thebibliography}{99} 
	
\bibitem{Abo-Shaeer01}
J. R. Abo-Shaeer, C. Raman, J. M. Vogels  and W. Ketterle W \textit{Science} \textbf{292} 476 (2001).

\bibitem{Coddington2003}
I. Coddington, P. Engels, V. Schweikhard, and E. A. Cornell,
Phys. Rev. Lett. \textbf{91}, 100402 (2003).


\bibitem{Donadello14}
S. Donadello {\it et al} \textit{Phys. Rev. Lett.} \textbf{113} 065302 (2014).

\bibitem{Wilson15}
K. E. Wilson, Z. L. Newman, J. D Lowney, and  B. P. Anderson, B. P.  Phys. Rev. A \textbf{91}, 023621 (2015).

\bibitem{Hodby01}
E. Hodby, G. Hechenblaikner, S. A. Hopkins, O. M. Maragò,
and C. J. Foot,  Phys. Rev. Lett. \textbf{88}, 010405 (2001).

\bibitem{Madison00}
K. W. Madison, F. Chevy, W. Wohlleben, and J. Dalibard,
Phys.	Rev. Lett. \textbf{84}, 806 (2000).

\bibitem{Haljan01}
P. C. Haljan, I. Coddington, P. Engels, and E. A. Cornell,
Phys. Rev. Lett. \textbf{87}, 210403 (2001).

\bibitem{Leanhardt02}
A. E. Leanhardt, A. Görlitz, A. P. Chikkatur, D. Kielpinski, Y.
Shin, D. E. Pritchard, and W. Ketterle,  Phys. Rev.
Lett. \textbf{89}, 190403 (2002).

\bibitem{Ripoll01}
J. J. García-Ripoll and V. M. Pérez-García
Phys. Rev. A \textbf{64}, 053611 (2001)

\bibitem{Aftalion2003a}
A. Aftalion, X. Blanc, and J. Dalibard
Phys. Rev. A \textbf{71}, 023611 (2005)

\bibitem{Engels2003a}
P. Engels, I. Coddington, P. C. Haljan, V. Schweikhard, and E. A. Cornell
Phys. Rev. Lett.\textbf{ 90}, 170405 (2003).

\bibitem{Fischer2003a}
U. R. Fischer and G. Baym, Phys. Rev. Lett. \textbf{90}, 140402 (2003).

\bibitem{Tkachenko}
V. K. Tkachenko \textit{Zh. Eksp. Teor. Fiz.} \textbf{49} 1875 (1965)

\bibitem{Tsubota02}
M. Tsubota, K. Kasamatsu and M. Ueda,  
Phys. Rev. A \textbf{65}, 023603 (2002).


\bibitem{Mueller}
E. J. Mueller, T. L.  Ho, M. Ueda and G. Baym   \textit{Phys. Rev.} A {\bf 74} 033612 (2006)

\bibitem{alon08} O.~E. Alon, A.~I. Streltsov, and L.~S. Cederbaum, Phys. Rev. A {\bf{77}}, 033613 (2008).

\bibitem{Lode20}
Lode, A.U.J.; Lévêque, C.; Madsen, L.B.; Streltsov, A.I.; Alon, O.E. 
Rev. Mod. Phys. 2020, \textbf{92}, 011001. (2020)

\bibitem{Lode_mctdhx}
R. Lin et. al.  Quantum Sci. Technol. \textbf{5} 024004 (2020)

\bibitem{chatterjee15} U.~R. Fischer, A.~U.~J. Lode, and B. Chatterjee, Phys. Rev. A \textbf{91}, 063621 (2015).

\bibitem{cao17} L. Cao, S.~I. Mistakidis, X. Deng, and P. Schmelcher, Chem. Phys. \textbf{482}, 303 (2017).


\bibitem{Mistakidis18a} S.~I. Mistakidis, G.~M. Koutentakis,  and P. Schmelcher, Chemical Physics, {\bf{509}}, 106 (2018).


\bibitem{Streltsova14} O.~I. Streltsova, O.~E. Alon, L.~S. Cederbaum, and A.~I. Streltsov, Phys. Rev. A \textbf{89}, 061602(R) (2014).



\bibitem{sakmann16} K. Sakmann and M. Kasevich, Nature Phys. {\bf 12}, 451 (2016).

\bibitem{lode17} A.~U.~J. Lode and C. Bruder, Phys. Rev. Lett. \textbf{118}, 013603 (2017).

\bibitem{tsatsos17} J.~H.~V. Nguyen, M.~C. Tsatsos, D. Luo, A.~U.~J. Lode, G.~D. Telles, V.~S. Bagnato, and R.~G. Hulet, Phys. Rev. X {\bf 9}, 011052 (2019).

\bibitem{chatterjee17a} B. Chatterjee and A.~U.~J. Lode, Phys. Rev. A \textbf{98}, 053624 (2018).

\bibitem{chatterjee17b} B. Chatterjee, M.~C. Tsatsos and A.~U.~J. Lode, New J. Phys. \textbf{21}, 033030 (2019).

\bibitem{chatterjee20}
B. Chatterjee, C. L\'ev\^eque, J. Schmiedmayer, and A.~U.~J. Lode,
Phys. Rev. Lett. \textbf{125}, 093602 (2020).

\bibitem{doganov} R.\,A. Doganov, S. Klaiman, O.\,E. Alon, A.\,I. Streltsov, and L.\,S. Cederbaum,, 
Phys. Rev. A {\bf 87}, 033631 (2013). 

\bibitem{Kramer} P. Kramer and M. Saraceno (ed)  \textit{Geometry of the Time-Dependent Variational Principle in Quantum
	Mechanics} (Berlin: Springer) (1981).


\end{thebibliography}
\end{document}